\documentclass[a4paper,11pt]{article}

\usepackage{booktabs}

\usepackage{lineno}
\usepackage[colorlinks=false]{hyperref}
\usepackage{bm}
\usepackage{graphicx}
\usepackage{amssymb}
\usepackage{amsmath}
\usepackage{tabularx}
\usepackage{natbib}
\usepackage{floatrow}
\usepackage{caption}
\usepackage{multirow}
\usepackage{amsmath}
\usepackage{amsfonts}
\usepackage{amssymb}
\usepackage{xcolor}
\usepackage{subfig}
\usepackage{graphicx,amsmath,bm}
\usepackage[binary-units=true]{siunitx}
\usepackage{array}
\usepackage{authblk}
\usepackage{rotating}
\usepackage{enumerate}
\usepackage{ragged2e}

\usepackage[left=15mm,right=15mm,top=1.5cm,bottom=1.5cm,includeheadfoot]{geometry}
\setcitestyle{square,numbers}
\begin{document}
	
	\title{Physics-Driven Neural Network for Solving Electromagnetic Inverse Scattering Problems}
	
	\author[1]{Yutong~Du}
	\author[1]{Zicheng~Liu}
	\author[2]{Bazargul~Matkerim}
	\author[1]{Changyou~Li}
	\author[1]{Yali~Zong}
	\author[1]{Bo~Qi}
	\author[3]{Jingwei~Kou}
	
	\affil[1]{\scriptsize Department of Electronic Engineering, School of Electronics and Information, Northwestern Polytechnical University, Xi'an 710029, China}
	\affil[2]{\scriptsize Department of Computer Science, Al-Farabi Kazakh National University, Almaty 050040, Kazakhstan}
	\affil[3]{\scriptsize The Advanced Optical Instrument Research Department, Xi’an Institute of Optics and Precision Mechanics, Chinese Academy of Sciences, Xi'an 710119, China}
	\maketitle
	
	\abstract{
		In recent years, deep learning-based methods have been proposed for solving inverse scattering problems (ISPs), but most of them heavily rely on data and suffer from limited generalization capabilities. In this paper, a new solving scheme is proposed where the solution is iteratively updated following the updating of the physics-driven neural network (PDNN), the hyperparameters of which are optimized by minimizing the loss function which incorporates the constraints from the collected scattered fields and the prior information about scatterers. Unlike data-driven neural network solvers, PDNN is trained only requiring the input of collected scattered fields and the computation of scattered fields corresponding to predicted solutions, thus avoids the generalization problem. Moreover, to accelerate the imaging efficiency, the subregion enclosing the scatterers is identified. Numerical and experimental results demonstrate that the proposed scheme has high reconstruction accuracy and strong stability, even when dealing with composite lossy scatterers.}
	
	\section{Introduction}
	Electromagnetic inverse scattering imaging\cite{chen2018computationalEMIS} technology plays a crucial role in various fields by reconstructing the geometric features (e.g., shape, size) and electromagnetic properties (e.g., relative permittivity, conductivity) of targets. Typical applications include non-destructive testing of composite fiber materials\cite{An2025TimeRev,Zhou2023PALS,An2024CompoMat}, contraband detection in airport security screening\cite{Wang2019SecScr,Ahmed2021MicroSec}, and detection of underground targets using ground penetration radar\cite{Esposito2024ExpVal,Dai2023GPR3D,Wang2022GPR2D}. The key to achieving quantitative analysis\cite{Adrian1988quant,A1988quant,Stefanov1990quant,Pedro2010quant,P2009quant} lies in addressing the inherent challenges of uniqueness, stability, and nonlinearity in inverse scattering problems (ISPs)\cite{chen2018computationalEMIS}. 
	
	The classical algorithms to solve the ISP challenges can be categorized into iterative and non-iterative methods. Non-iterative methods include the Born approximation (BA) method\cite{habashyi1993BA,gao2006BA}, backpropagation (BP) method\cite{devaney1982BP,tsili1998BP}, and Rytov approximation (RA) method\cite{devaney1981RA,alon1993RA,slaney1984RA}. These methods rely on linearization assumptions and cannot fully consider multiple scattering effects within scatterers, making them only suitable for low-contrast scatterers. Iterative methods, such as Born iterative method (BIM)\cite{wang1989BIM,sung1999BIM}, distorted Born iterative method (DBIM)\cite{chew1990DBIM,haddadin1995DBIM}, contrast source inversion (CSI) method\cite{peter1997CSI,richard2001CSI} and subspace optimization method (SOM)\cite{chen2010SOM,chen_2010SOM,pan2011SOM} iteratively searching for the exact solution and the iteration scheme leads to the possible high-computational cost, despite the effects from initial solution and convergence rate. 
	
	Deep learning techniques have been applied for solving ISPs. Wei et al. proposed three U-Net-based inversion schemes: the direct inversion scheme (DIS), backpropagation scheme (BPS), and dominant current scheme (DCS). These schemes achieved computation time of less than 1 second outperforming classical iterative methods\cite{wei2018BPS}. Li et al. introduced DeepNIS to address nonlinear electromagnetic ISPs for large-scale and high-contrast objects, which demonstrates remarkable improvements in both image quality and computational efficiency compared to conventional nonlinear inverse scattering methods\cite{Li2018DeepNIS}. The neural network solvers learn the data features, based on which the desired solution can be accurately represented, but often ignore the inherent physical laws of the imaging process. As a result, data-driven deep neural network (DNN) solvers always have challenges in interpretability and generalizability.
	
	In recent years, researchers have made efforts to integrate physical knowledge into DNN solvers. Wei et al. proposed the induced current learning method (ICLM), which the contrast source is reconstructed to solve ISPs and a cascaded end-to-end neural network architecture is applied to decrease the nonlinearity of objective function \cite{wei2019PhaNN}. Liu et al. improved noise robustness by incorporating the loss term with respect to the near scattered fields and enhanced feature learning ability by putting constraints about the induced currents within domain of interests (DOI)\cite{Liu2022PhaGuideNN}. In \cite{Liu2022SOMnet}, the SOM-Net ISP solver is proposed where induced currents and contrast distribution are alternatively updated in an iteration solving scheme with cascaded sub-networks. Shan et al. developed NeuralBIM where the training process of the neural network solver is guided by physics-embedded objective function derived from the governing equations of ISPs \cite{Tao2023NNBIM}. While efforts have made on the ISP solver through the design of loss functions and neural network architecture, the solvers are still data driven and the generalization ability is always a challenge which needs to be tested comprehensively.
	
	In this paper, we proposes an ISP solving scheme based on a physics-driven neural network (PDNN), which is trained only making use the scattered fields collected by receivers and the imaging physical laws. Following the classical iterative methods, the solution is iteratively retrieved while at each iteration the updated solution is predicted by the neural network, the hyperparameters of which are optimized by minimizing the loss function composed of scattered-fields data discrepancy term and the two terms imposing prior knowledge about scatterers to reduce the solution domain. The key contributions of the presented work are summarized below: 
	
	(1) Apply a physics-driven neural network in the iterative solving process to establish the mapping relation between the collected scattered fields to the scatterer profile. Unlike traditional methods (e.g., gradient descent, Newton's method) that rely on local approximations and can converge to local minima in highly nonlinear, non-convex problems, the proposed PDNN utilizes a multilayer neural network architecture and nonlinear activation functions (ReLU/LeakyReLU) to enable the solver learning the complex mapping relation and reaching the global optima. 
	
	(2) Leverage the governing physical laws to guide the updating of hyperparameters of the neural network so that the obtained solution is physics consistent. Specifically, through penalizing the discrepancy between the scattered fields collected by receivers and the ones computed corresponding to the predicted solution, the imaging physics have been fully exploited and obeyed in the reconstruction process. 
	
	(3) Enhance the imaging efficiency by identifying the valid imaging region. The computation cost is proportional to the number of grids in the domain of interests (DOI). In the paper, based on the solution from U-Net solver, after the morphological procedures including thresholding, closing and dilation, the subregion enclosing the true scatterers is identified and used to significantly reduce the computational burden.
	
	(4) Design a new loss function imposing constraints on the lower bound of the relative permittivity and the piece-wise homogeneity of the permittivity distribution. The additional constraints can effectively reduce the solution space and improve the reconstruction accuracy and stability. 
	
	(5) Non data-driven learning scheme avoids the generalization problem which is often accounted by data-based solving algorithms. Since the solver is trained independently w.r.t. each imaging case and no features from the training data are relied, the PDNN solver enjoys excellent imaging performances for all kinds of scatterers which are analyzed in Section \ref{sec:NumeSimu}. 
	
	This paper is organized as follows. In Section \ref{sec:formulateISPs}, the concerned ISP problems and the involved physical laws are illustrated. Section \ref{sec:PINNSsolvers} gives the details of the proposed solving scheme based on a neural network with the introduction of procedures to reduce the imaging region and the metric quantifying the imaging accuracy. The imaging performances are analyzed based on numerical results in Section~\ref{sec:NumeSimu} and experimentally verified in Section \ref{sec:forExp}. Conclusions are made in Section \ref{sec:conclusions}.

	\section{Formulation of ISPs}
	\label{sec:formulateISPs}
	\begin{figure}[!t]
		\centering
		\includegraphics[width = 0.4\linewidth]{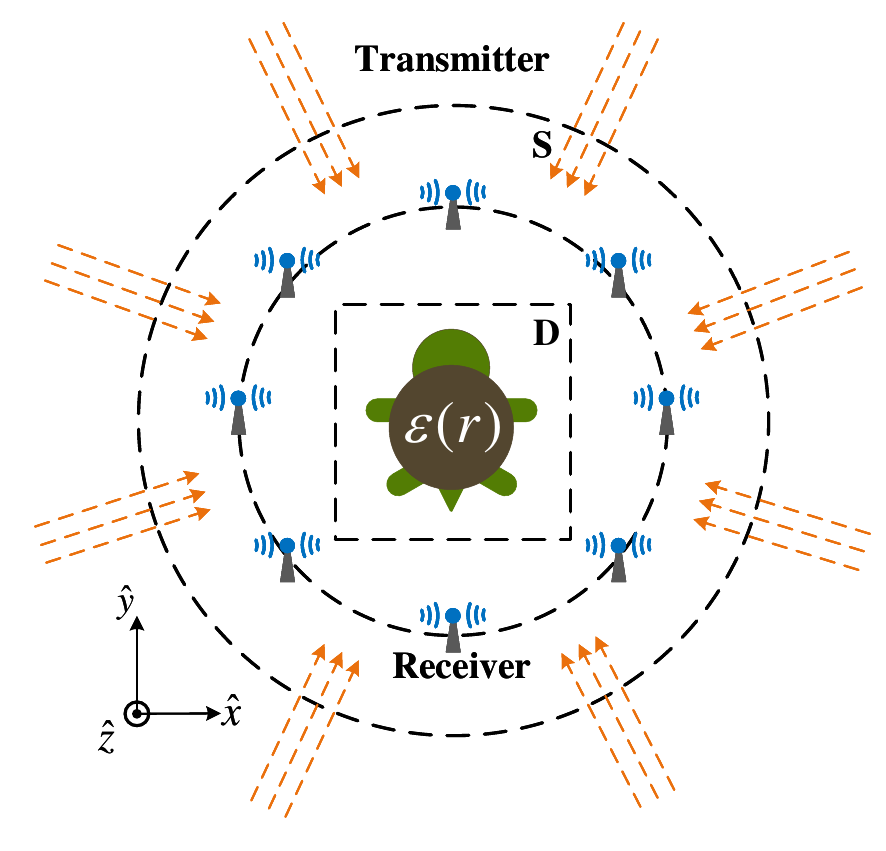}
		\caption{The diagram for the concerned imaging configurations with multiple transmitters and receivers.}
		\label{fig:ISPs}
	\end{figure}
	The concerned imaging system is sketched in Figure.~\ref{fig:ISPs}. Here, two-dimensional ISPs are considered by detecting the domain of interests (DOI) with transverse magnetic (TM) waves and scatterers in DOI are homogeneous along the $z$-direction. ${N_{i}}$ transmitters alternatively illumate the DOI, and the induced current sources radiate scattered fields, which are collected by ${N_{s}}$ receivers.
	
	The forward and inverse problem is governed by the state equation and the data equation\cite{chen2018computationalEMIS}, which are expressed as 
	\begin{equation}
		\mathbf{E}^\text{tot}(\mathbf{r}) = \mathbf{E}^\text{inc}(\mathbf{r}) + k_0^2\int_\text{DOI}g(\mathbf{r},\mathbf{r}^\prime)\mathbf{J}(\mathbf{r}^\prime)d\mathbf{r}^\prime, \mathbf{r}\in\text{DOI},
		\label{eq:stateEqu}
	\end{equation}
	and 
	\begin{equation}
		\mathbf{E}^\text{sca}(\mathbf{r}) = k_0^2\int_\text{DOI}g(\mathbf{r},\mathbf{r}^\prime)\mathbf{J}(\mathbf{r}^\prime)d\mathbf{r}^\prime, \mathbf{r}\in\text{S},
		\label{eq:dataEqu}
	\end{equation} 
	respectively. $\mathbf{E}^\text{tot}(\mathbf{r})$ denotes the $z$ component of total electric field at the observation position $\mathbf{r}$, while $\mathbf{E}^\text{inc}$ and $\mathbf{E}^\text{sca}$ stand for the incident and the scattered one. $k_0$ is the wavenumber of background medium, $g$ is the two-dimensional scalar Green's function, and the induced current density $\mathbf{J}(\mathbf{r}^\prime)=(\epsilon_r(\mathbf{r}^\prime)-1)\mathbf{E}^\text{tot}(\mathbf{r}^\prime)$, where $\epsilon_r$ is the relative permittivity. ``S" denotes the region where scattered fields are collected. 
	
	Denoting $\mathbf{G}_\text{D}$ and $\mathbf{G}_\text{S}$ as the integral operator for the state equation and the data equation, respectively, \eqref{eq:stateEqu} and \eqref{eq:dataEqu} can be rewritten as
	\begin{equation}
		\mathbf{E}^\text{tot}(\mathbf{r}) = \mathbf{E}^\text{inc}(\mathbf{r}) + \mathbf{G}_\text{D}({\boldsymbol{\epsilon_r}}-1)\mathbf{E}^\text{tot}, \mathbf{r}\in\text{DOI},
		\label{eq:GreenOperatorStateEqu}
	\end{equation}
	\begin{equation}
		\mathbf{E}^\text{sca}(\mathbf{r}) = \mathbf{G}_\text{S}({\boldsymbol{\epsilon_r}}-1)\mathbf{E}^\text{tot}, \mathbf{r}\in\text{S}.
		\label{eq:GreenOperatorDataEqu}
	\end{equation} 
	%Then, the scattered field is solved by \cite{Ney1985MoM,Gibson2021MoM,LAKHTAKIA1992MoM}
	%\begin{equation}
	%	\mathbf{E}^\text{sca} = \mathbf{G}_\text{S}\left({\boldsymbol{\epsilon_r}}-1\right)\left(\mathbf{I}-\mathbf{G}_\text{D}\left({\boldsymbol{\epsilon_r}}-1\right)\right)^\mathbf{-1}\mathbf{E}^\text{inc},
	%	\label{eq:FinalDataEqu}
	%\end{equation}
	
	The inverse problem aims to reconstruct the distribution of $\boldsymbol{\epsilon_r}$ within the DOI based on the measured scattered fields $\mathbf{E}^\text{sca}_\text{mea}$. Regularization techniques\cite{Oliveri2017Regular} are widely used to stabilize ISP solutions by incorporating additional constraints or prior information, where the objective function can be formulated as
	\begin{equation}
		L({\hat{\boldsymbol{\epsilon_r}}}) =||{\mathbf{E}}^\text{sca}_\text{mea}-\hat{\mathbf{E}}^\text{sca}_\text{mea}({\hat{\boldsymbol{\epsilon_r}}})||_1 +  g({\hat{\boldsymbol{\epsilon_r}}}),
		\label{eq:optimizationEq}
	\end{equation} 
	where the first term imposes data-discrepancies constraint between the collected scattered fields and the scattered electric fields corresponding to the solution $\hat{\boldsymbol{\epsilon_r}}$, \emph{i.e.}, $\hat{\mathbf{E}}^\text{sca}_\text{mea}$, and the second term is the regularization term.
	
	\section{Physics-driven neural network solvers}
	\label{sec:PINNSsolvers}
	A physics-driven neural network (PDNN) is proposed to retrieve ISP solutions with a loss function  applying comprehensive constraints to guide the neural network solver generating solutions fitting the inherent physical laws.
	
	\subsection{Inversion scheme}
	\label{subsec:nnArch}
	\begin{figure*}[!t]
		\centering
		\includegraphics[width = \linewidth]{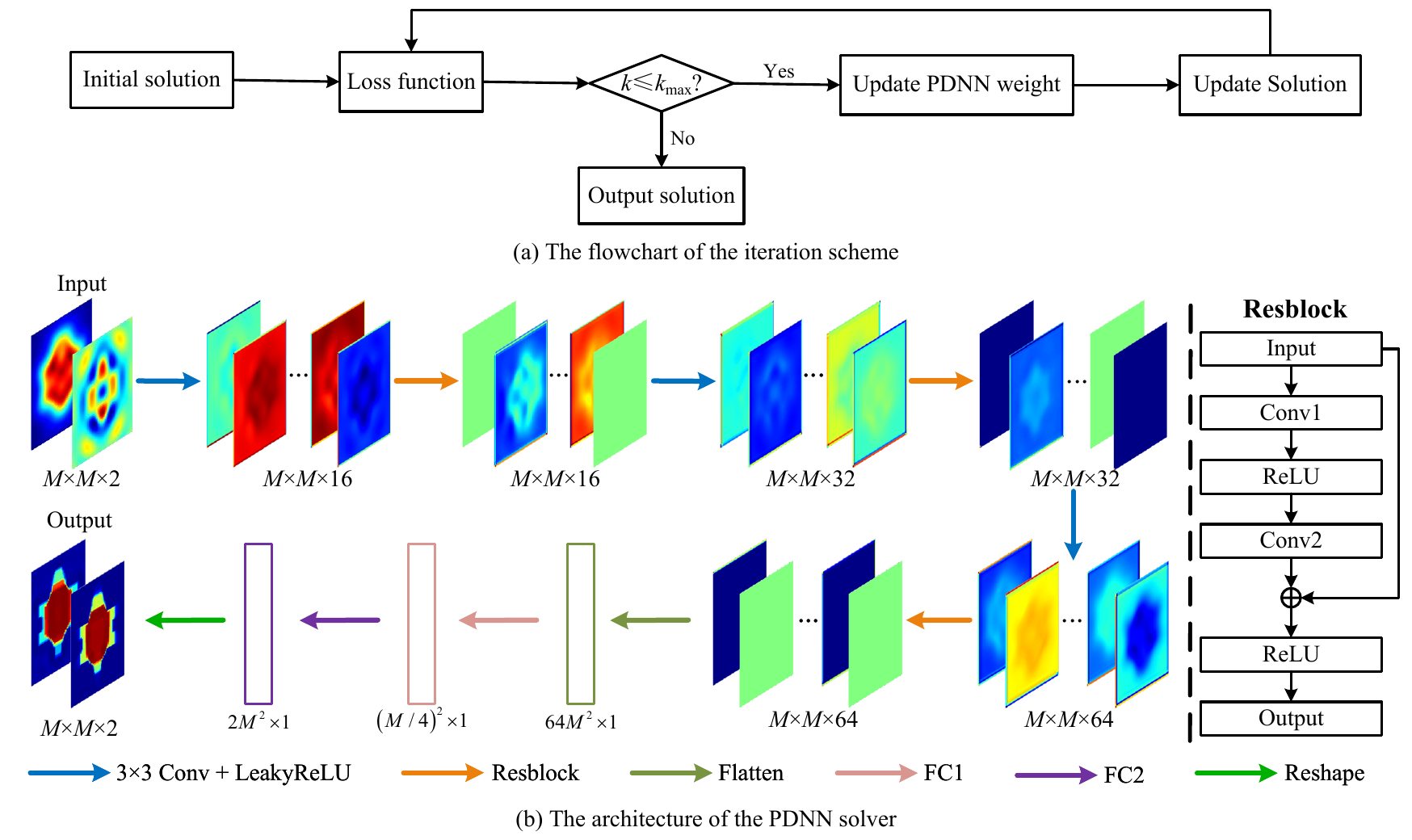}
		\caption{Sketch of the proposed inversion scheme based on physics-driven neural network (PDNN).}
		\label{fig:PDNN}
	\end{figure*}
	
	The proposed inversion scheme is sketched in Figure.~\ref{fig:PDNN}(a). Starting from an initial solution, the solution is iteratively updated with the PDNN and the iteration process is ended until the stopping criteria is satisfied, \emph{i.e.}, the inversion scheme is same with the classical iterative ISP solvers except that the solution is updated by the neural network.  
	
	However, unlike data-driven neural network solvers which are obtained based on hundreds or thousands of training data samples, the proposed inversion scheme is based on the inherent imaging physical laws that are governed by the data equation and the state equation through the designed loss function (explained below). Thus, the neural network solver w.r.t. different testing scenarios need to be built independently and the solver has no limit on the generalization ability which is usually lacked for data-driven neural network solvers.   
	
	The applied PDNN architecture is sketched in Figure.~\ref{fig:PDNN}(b). The architecure consists of three convolutional layers, three residual blocks, and two fully connected layers, where convolutional layers learn data features, residual connections effectively mitigate vanishing gradient problems, and fully connected layers yield the predicted contrast distribution based on the learned features. ReLU and LeakyReLU (negative slope being 0.01) activation functions are applied to enable the network learning nonlinear relations of ISPs. Both the input and the response are composed of two channels which stand for the real and the imaginary part of contrast so that the solver based on the real-valued neural network is compatible to lossy dielectric scatterers. 
	
	\subsection{Loss function}
	\label{subsec:lossFunc}
	The loss function applied by the PDNN solver is a summation of three items, \emph{i.e.},
	\begin{equation}
		\text{Loss} = L^{\text{Data}}+\alpha L^{\text{Bound}}+\beta L^{\text{TV}},
		\label{LossFunc}
	\end{equation}
	and the three items are defined as 
	\begin{subequations}
		\begin{align}
			\label{eq:LData}
			&L^{\text{Data}}=||{\mathbf{E}}^\text{sca}_\text{mea}-\hat{\mathbf{E}}^\text{sca}({\hat{\boldsymbol{\epsilon}}_r})||_1,\\
			\label{eq:LreEp}
			&L^{\text{Bound}}=||\text{ReLU}(1-\mathbf{Re\{\hat{\boldsymbol{\epsilon}}_r\}})||_1,\\
			\label{eq:LTV}
			&L^{\text{TV}}=\sum_{i,j}\left[(\boldsymbol{\epsilon}_{i,j-1}-\boldsymbol{\epsilon}_{i,j})^2+(\boldsymbol{\epsilon}_{i+1,j}-\boldsymbol{\epsilon}_{i,j}\right)^2].
		\end{align}
	\end{subequations}
	$L^{\text{Data}}$ quantifies the residual between the measured scattered fields $\mathbf{E}^\text{sca}_\text{mea}$ and the scattered fields corresponding to the predicted solution $\hat{\boldsymbol{\epsilon}}_r$. $L^{\text{Bound}}$ is to impose the constraint on the lower bound of the real part of relative permittivity which is set as 1 here. $L^{\text{TV}}$ is the total-variation regularization term \cite{Leonid1992TV}, imposing smoothness on the desired solution. $\alpha$ and $\beta$ are the hyperparameters trading off the contributions from different terms. Their optimal setting is discussed in Section~\ref{subsec:hyperparameter}.
	
	From \eqref{eq:LData}, one see that the scattered fields corresponding to the update solution, $\hat{\mathbf{E}}^\text{sca}({\hat{\boldsymbol{\epsilon}}_r})$, need to be retrieved. Method of moments (MoM)\cite{Ney1985MoM,Gibson2021MoM,LAKHTAKIA1992MoM} is applied to compute $\hat{\mathbf{E}}^\text{sca}({\hat{\boldsymbol{\epsilon}}_r})$ based on data equation and state equation. However, the computational cost can be high when the number of iterations is large. 
	
	\subsection{Scatter subregion identification}
	\label{subSec:reduceDOI}	
	The computational complexity of MoM is proportional to the number of grids in domain of interests. To reduce the computational cost, the number of grids is reduced through cropping the imaging domain based on an initial estimation of contrast distribution. 
	
	Based on measured scattered fields, back propagation (BP) method can be used to generate a fast solution of scatter profile. However, BP results are often blurred or corrupted by background artifacts which lead to challenges in distinguishing the scatter edges. The classical U-Net-based neural network solver presented in \cite{wei2018BPS} is applied to have high-quality estimate. The neural network solver is data-driven. After training offline, the solver is able to generate solution online efficiently. 
	
	To identify the scatter region and the background, the statistical character of background contrast values is evaluated. Resorting all values of the reconstructed contrast distribution in descent order, the mean and the standard deviation of the first 50\% values are computed and denoted as $\mu$ and $\sigma$, respectively. Then, the scatter positions can be identified by comparing the contrast value with the threshold $\mu+\delta\sigma$ and the position indices compose the binary map \cite{gonzalez2017Threshold} defined by 
	\begin{equation}
		\text{B}^\text{Thresholding}(i,j) = \left\{
		\begin{array}{cl}
			1, & \text{if} \, \chi(i,j) \ge \mu+\delta\sigma \\
			0, & \text{otherwise}
		\end{array}
		\right.
	\end{equation}
	However, due to the data-driven property, the U-Net-based solver may has limitations in generalization ability, \emph{i.e.}, artifacts can happen and the scatter edges maybe not accurately reconstructed. As a result, the obtained $\text{B}^\text{Thresholding}$ may not accurately indicate the true scatter region. 
	
	To avoid such problems, the value of $\delta$ is conservatively set as $3$ to include more grids as scatter region. Considering the imaging challenges of adjacent scatterers, a morphological closing operation with a structuring element $\text{S}$ is applied to $\text{B}^\text{Thresholding}$ to connect nearby objects, \emph{i.e.},
	\begin{equation}
		\text{B}^\text{closing}=(\text{B}^\text{Thresholding}\oplus\text{S})\ominus\text{S}
		\label{closed}
	\end{equation}
	where $\oplus$ and $\ominus$ represent morphological dilation and erosion operators respectively. Here, the structuring element $\text{S}$ is chosen as a disk with radius $r$.
	
	To further ensure complete coverage of potential scatterers, $\text{B}^\text{closing}$ is dilated using the same disk-shaped structuring element $\text{S}$, enlarging the detected regions as
	\begin{equation}
		\text{B}^\text{Dilation}=\text{B}^\text{closing}\oplus\text{S}.
		\label{Expanded}
	\end{equation}
	
	\begin{figure*}[!t]
		\centering
		\includegraphics[width = .9\linewidth]{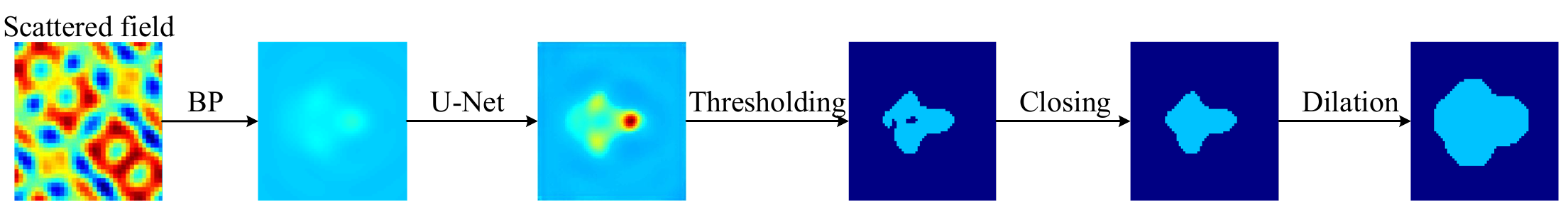}
		\caption{Identify the subregion enclosing scatterers based on the solution from U-Net solver \cite{Liu2022PhaGuideNN} after morphological operations.}
		\label{fig:process}
	\end{figure*}
	
	An example running the above procedures to have $\text{B}^\text{Dilation}$ is presented in Figure.~\ref{fig:process}. The applied strategy to determine the imaging region is to ensure that all potential scatterer grids are included, even at the cost of mistakenly including some background ones. Restricting the imaging region as $\text{B}^\text{Dilation}$, the computational cost to retrieve $\hat{\mathbf{E}}^\text{sca}$ can be significantly reduced without sacrificing the imaging accuracy, as presented in Section \ref{subsec:enhancedPDNN}.
	
	\subsection{Training settings and evaluation indicator}
	\label{subsec:trainSet}
	In this paper, all ISP solvers are trained on a workstation equipped with 128 GB RAM, a 3.2 GHz i9 CPU and an NVIDIA GeForce RTX 4090 GPU. The neural network parameters are optimized using the Adam algorithm with the learning rate of 1$\times$10$^{-3}$. The number of training iterations is discussed in Section \ref{subsec:Iterations}.
	
	The discrepancy between the reconstructed relative-permittivity distribution and the ground truth is quantified by the relative error defined by
	\begin{equation}
		\delta = {\frac{||\mathbf{Re\{\hat{\boldsymbol{\epsilon}}_r\}}-\mathbf{Re\{{\boldsymbol{\epsilon_r}}\}}||_1+||\mathbf{Im\{\hat{\boldsymbol{\epsilon}}_r\}}-\mathbf{Im\{{\boldsymbol{\epsilon_r}}\}}||_1}{||\mathbf{Re\{{\boldsymbol{\epsilon_r}}\}}||_1+||\mathbf{Im\{{\boldsymbol{\epsilon_r}}\}}||_1}}
	\end{equation}
	
	\section{Numerical Analysis}
	\label{sec:NumeSimu}
	To analyze the parameter of the PDNN scheme and evaluate the imaging performances, tests are performed based on simulated data when the DOI is sized by 0.15m$\times$0.15m and discretized into $M\times M$ grids, where $M$ = 64. MoM is employed to compute the scattered fields due to $36$ transmitters and $36$ receivers which are uniformly distributed on the circle sharing the same center with DOI and with radius $20\lambda$, $\lambda$ being the wavelength corresponding to the wave frequency 4 GHz. 
	
	The classical U-Net model is trained based on 1000 digit-like profiles with relative permittivity values ranging from 1 to 5. The Euclidean distance $||\hat{\boldsymbol{\chi}}-\boldsymbol{\chi}||^2$ is used as the loss function, where $\hat{\boldsymbol{\chi}}$ and $\boldsymbol{\chi}$ denote the predicted contrast distribution and the ground truth, respectively. The training parameters are same with the reference \cite{Liu2022PhaGuideNN}. 
	
	To test the imaging performances comprehensively, four representative scatter profiles designed: the square profile with side length of $0.5\lambda$ to verify whether PDNN can accurately reconstruct square corners, the two-circles profile with the same radius of $0.25\lambda$ and centered at $(\pm4\lambda, 0)$ to evaluate the ability of PDNN to resolve scatterers close to each other, the ring profile characterized by the inner radius $0.5\lambda$ and the outer one $0.25\lambda$ as a challenge of recovering the interior edge of scatterers, and the Austria profile which combines circular and annulus scatterers and suffers from complex multiple scattering effects. The Austria profile consists of two circles centered at $(\pm0.4\lambda$, $0.65\lambda)$ and with the identical radii of $0.2\lambda$, and an annulus profile centered at $(0, -0.25\lambda)$ and with outer radii of $0.6\lambda$ and inner radii $0.3\lambda$.
	
	\subsection{Influence of hyperparameters in loss function}
	\label{subsec:hyperparameter}
		
	\begin{figure*}[!ht]
		\centering
		\includegraphics[width = 0.75\linewidth]{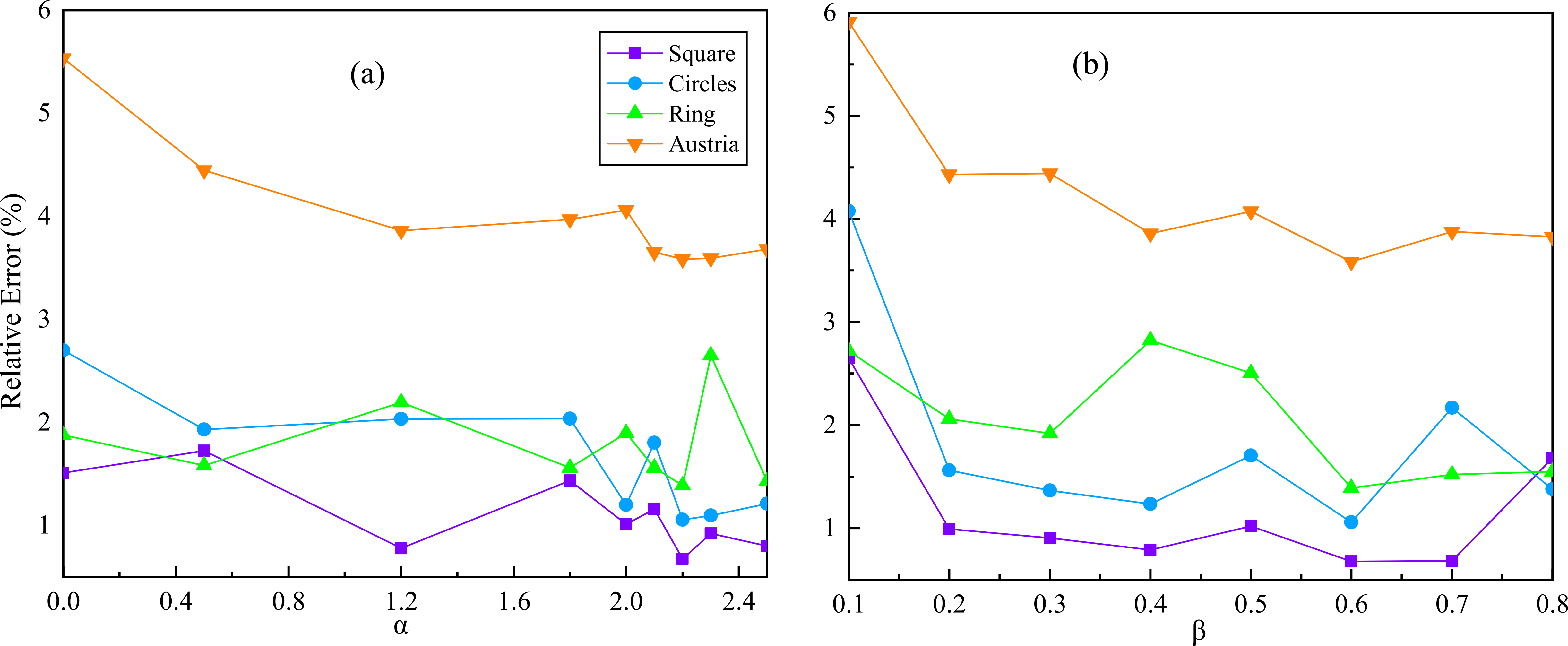}
		\caption{Impact of hyperparameters in the loss function on the prediction error for the four representative profiles: square, circles, ring, and Austria when fixing (a) $\beta = 0.6$, or (b) $\alpha=2.2$.}                   
		\label{fig:alpha_beta}
	\end{figure*}
	
	\begin{figure*}[!ht]
		\centering
		\includegraphics[width = .9\linewidth]{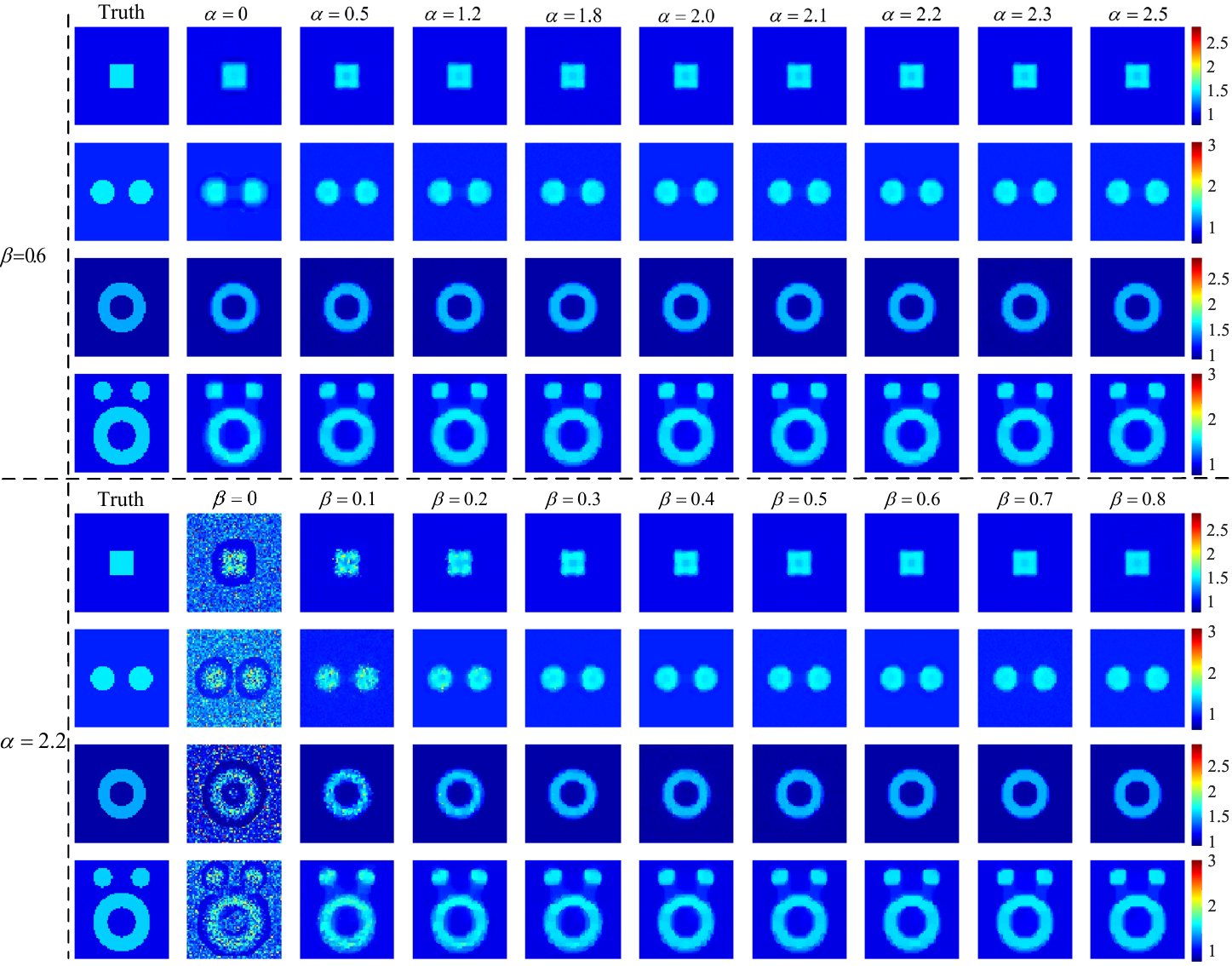}
		\caption{Imaging results for the four representative profiles with different values of $\alpha$ and $\beta$.}   
		\label{fig:findHyper}
	\end{figure*}
	
	The optimal value for the hyperparameter $\alpha$ and $\beta$ in the loss function is determined with a systematic testing. Fixing $\beta$ as 0.6, which is found optimal as shown below, and varying $\alpha$ from 0 to 2.5, the curve of prediction relative error $\delta$ is plotted in Figure.~\ref{fig:alpha_beta}(a). As seen, although $\alpha$ has impacts on the reconstruction accuracy, the relative error is below 6\% for $\alpha$ in the studied range, \emph{i.e.}, the imaging performance is not sensitive to the value of $\alpha$. However, from the observation of the cases with smallest relative error, $\alpha = 2.2$ is chosen as the optimal value. For better comparisons in vision, Figure.~\ref{fig:findHyper} gives the corresponding imaging results with different values of $\alpha$, which confirm the above conclusions regarding the selection of hyperparameter $\alpha$.
	
	Then, setting $\alpha$ as 2.2, the studies on the effects from the value of $\beta$ are performed by varying $\beta$ from 0 to 0.8 and computing the reconstruction relative error $\delta$. The results are plotted in Figure.~\ref{fig:alpha_beta}(b). With the increase of $\beta$, $\delta$ tends to decrease until $\beta$ reaches 0.6. From the corresponding imaging results in Figure.~\ref{fig:findHyper}, one sees that imposing smoothness regularization is effective to suppress the background artifacts and recover the homogeneous property of the scatterers. One can also find that the imaging performance is not sensitive to the value of $\beta$ when $\beta\ge 0.4$. $0.6$ is chosen as the optimal value considering the value of $\delta$ for the four representative profiles.
	
	Based on the above analysis, $\alpha=2.2$ and $\beta=0.6$ are considered optimal and applied in the following analysis.
			
	\begin{figure*}[!ht]
		\centering
		\includegraphics[width = .9\linewidth]{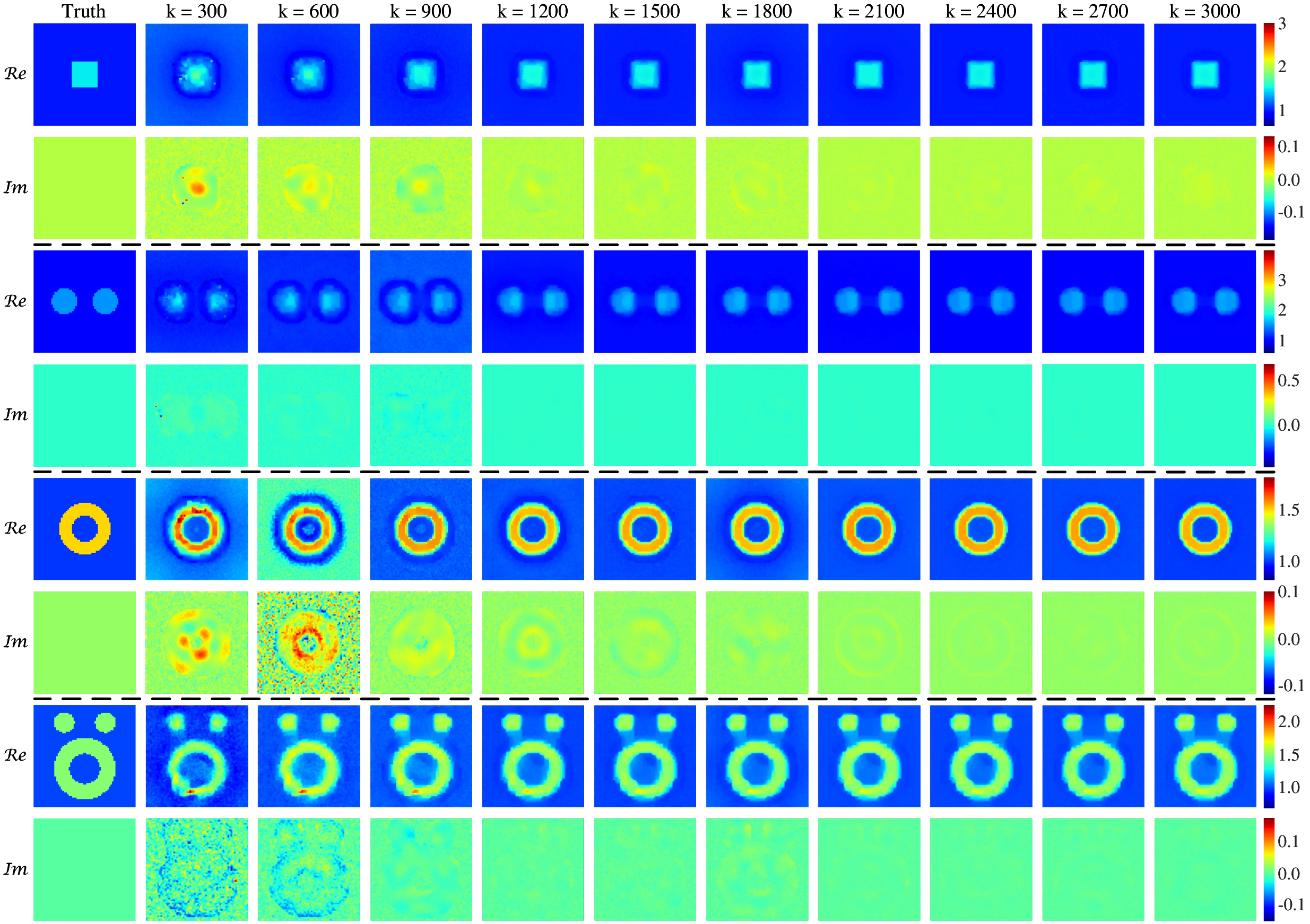}
		\caption{The imaging results of the four representative profiles after every 300 iterations when the maximum iteration number set as 3000.}
		\label{fig:Iteration}
	\end{figure*}
	
	\begin{figure}[!ht]
		\centering
		\includegraphics[width = 0.5\linewidth]{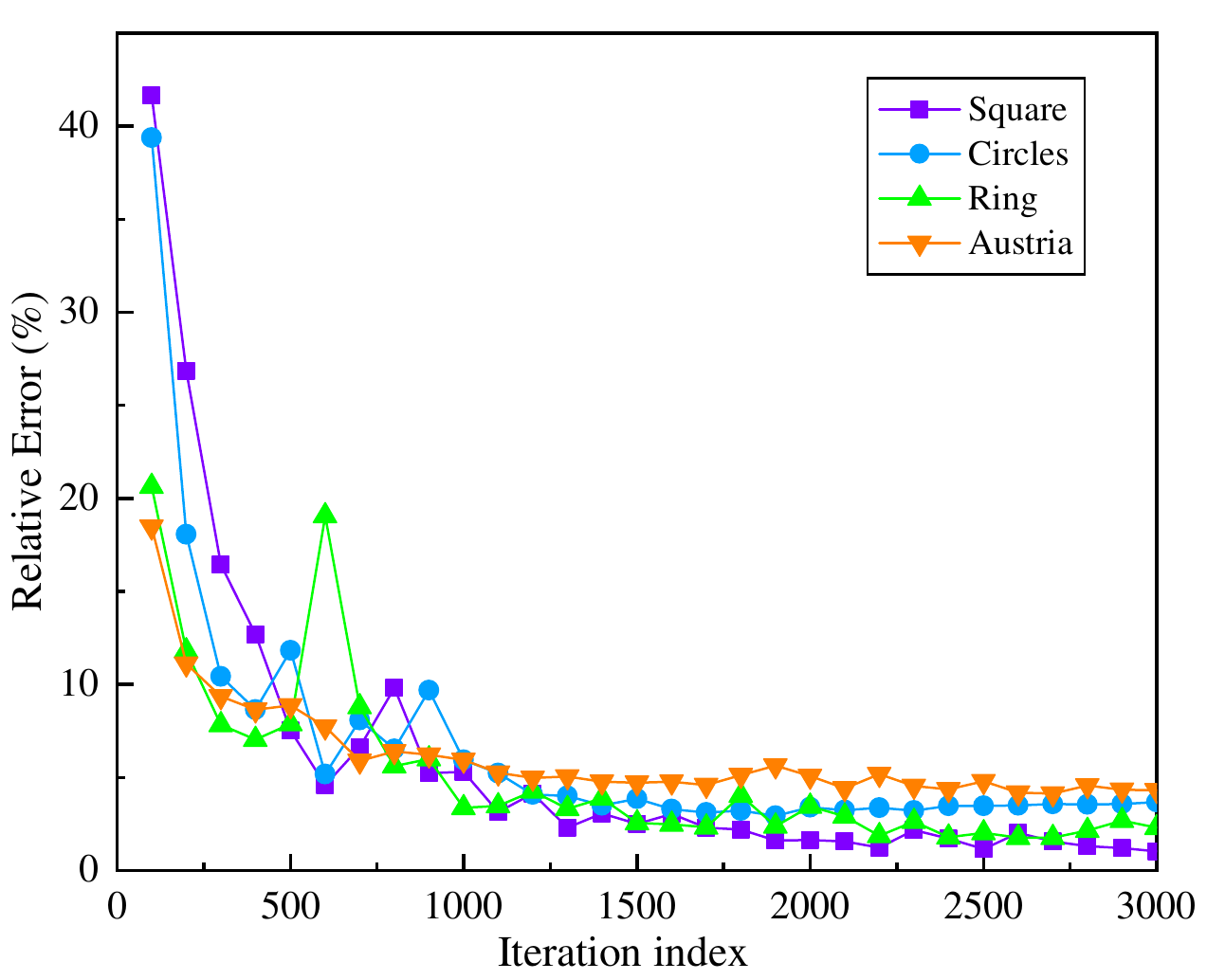}
		\caption{As iteration index increases, the imaging results show convergence trend with decreasing prediction error.}
		\label{fig:IterationCurve}
	\end{figure}
	
	\subsection{Influence of iteration number}
	\label{subsec:Iterations}	
	The convergence of the obtained solution w.r.t. the iteration number $k$ is analyzed here. The imaging results in Figure.~\ref{fig:Iteration} present the evolvement of relative-permittivity distribution every 300 iterations. When the iteration number $k=300$, the shape of the four profiles can be identified but strong artifacts appear. As $k$ increases further, both the real part and imaginary part of the relative permittivity converge to the ground truth. Quantifying the imaging accuracy by the defined relative error, the convergence phenomenon can also be observed from the curve of relative error given in Figure.~\ref{fig:IterationCurve}. The convergences happen for all the four representative profiles. The relative error has a sharp reduction when $k<1000$ and tends to be converged when $k$ increases further. 
				
	\begin{figure*}[!th]
		\centering
		\includegraphics[width = \linewidth]{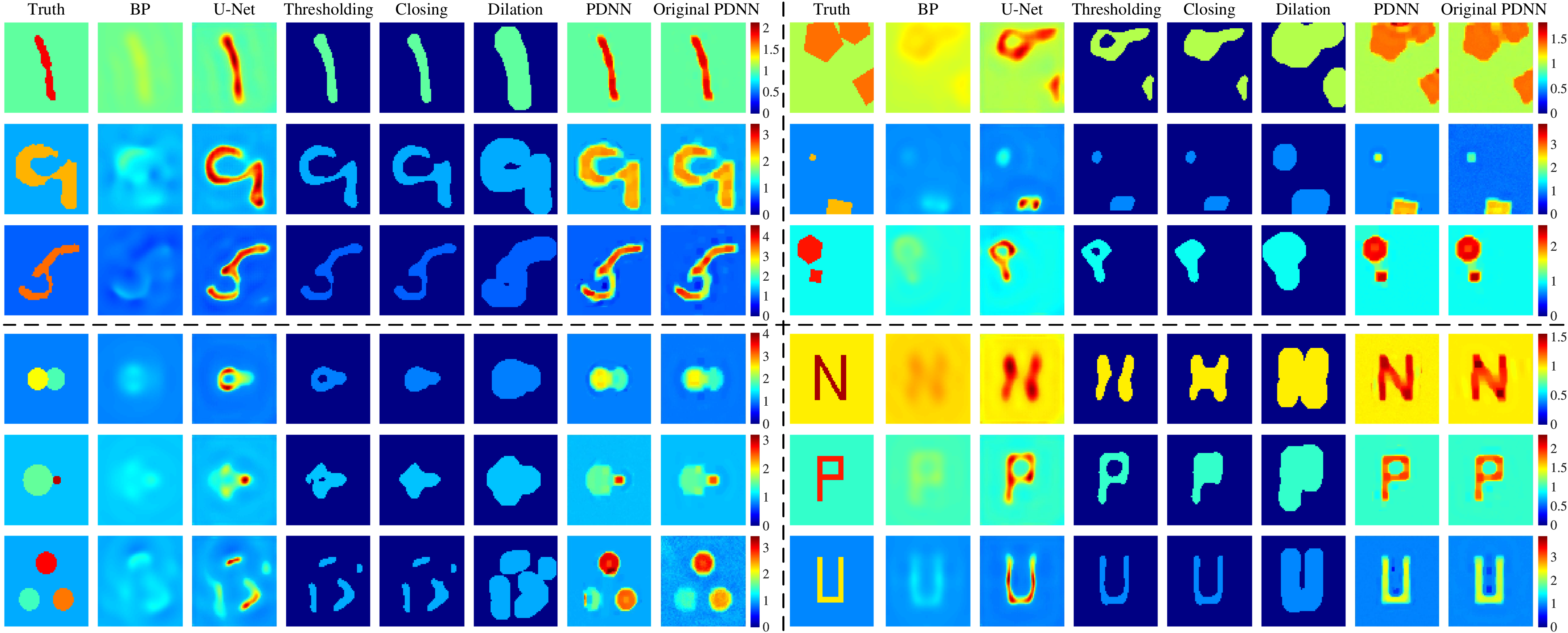}
		\caption{Comparisons of imaging results from the classical ISP solvers BP, U-Net and the proposed PDNN solvers including the original one and the one with reduced imaging region.}
		\label{fig:enhancedPDNN}
	\end{figure*}
	
	\begin{table*}[!ht]
		\centering
		\resizebox{1.0\columnwidth}{!}{
		\caption{Reduction of grids number and runtime after applying the imaging subregion identification strategy}
		\label{tab:OrEnhancedPDNN}
		\resizebox{2.0\columnwidth}{!}{
			\begin{tabular}{l|c c| c c| c c|c c}
				\toprule
				\textbf{} & \multicolumn{2}{c|}{\textbf{Digit-like}} & \multicolumn{2}{c|}{\textbf{Polygon-like}} & \multicolumn{2}{c|}{\textbf{Multiple circles}} & \multicolumn{2}{c}{\textbf{Letter}} \\ 
				\midrule
				\textbf{Indicators} & \textbf{Grids number} & \textbf{Runtime} & \textbf{Grids number} & \textbf{Runtime} & \textbf{Grids number} & \textbf{Runtime} & \textbf{Grids number} & \textbf{Runtime} \\ 
				\midrule
				Case 1   & $4096\to992$ & $362s\to54s$ &$4096\to1717$ & $362s\to74s$ & $4096\to831$ & $362s\to48s$ & $4096\to1482$ & $362s\to66s$          \\ 
				Case 2   & $4096\to1949$ & $362s\to85s$ &$4096\to777$ & $362s\to49s$ & $4096\to988$ & $362s\to52s$ & $4096\to1210$ & $362s\to55s$       \\ 
				Case 3   & $4096\to1467$ & $362s\to65s$ &$4096\to945$ & $362s\to51s$ & $4096\to1623$ & $362s\to71s$ & $4096\to1255$ & $362s\to58s$       \\ 
				\bottomrule
			\end{tabular}
		}
		}
	\end{table*}

	\subsection{Runtime acceleration of PDNN}
	\label{subsec:enhancedPDNN}	
	As mentioned in Section \ref{subSec:reduceDOI}, the imaging region is reduced to lighten the computation burden due to the updating of scattered fields required by the computation of loss function. The acceleration effects are tested based on four representative groups of profiles: digit-like, polygon-like, multiple circles and letter profiles, and each group includes three representative examples, the image of which is shown in Figure.~\ref{fig:enhancedPDNN}.
	
	To indicate the acceleration effects, the number of grids and the runtime before (denoted by ``Original PDNN") and after (denoted by ``PDNN") the imaging region reduction procedures are summarized in Table~\ref{tab:OrEnhancedPDNN} where the iteration number is fixed as 3000. As observed, the number of grids is at least halved, even more than 80\% reduced, and the runtime is substantially decreased compared to the original PDNN. 
	
	From Figure.~\ref{fig:enhancedPDNN}, one see that for digit-like profiles, U-Net, original PDNN and PDNN can accurately reconstruct the geometric features. Since the testing profile is the same kind with the training data, U-Net shows superior performances in terms of geometry reconstruction accuracy. In contrast, PDNN and the original one maybe inferior to U-Net in reconstructing the digit ``5” profile, but have advantages in relative-permittivity reconstruction accuracy. Comparing PDNN and the original one, background artifacts are more successfully suppressed with PDNN since a part of background region is identified and not in the imaging region during the imaging process. 
	
	For polygon-like profiles, U-Net can well reconstruct small scatterers, but is lacking the ability of recovering the shape of big scatterers. More specifically, hollow artifacts appear at the central area of the scatterers with big size. As a result, after the thresholding step, a binary map which cannot cover the whole scatter region is obtained. However, after closing and the dilation operations, the imaging region has been expanded such that all scatterers have been enclosed. Then, the PDNN and the original version both can reconstruct the scatter profile accurately with significant advantages over U-Net.
	
	For the multiple circles profiles, U-Net performs poorly considering the shape of circular scatterers is distorted and the piece-wise homogeneity cannot identified. The U-Net model is trained based on the dataset composed of digit-like sample and the learned features may not be proper to represent circular-shape scatterers. This phenomenon reveals that data-driven ISP models maybe lacking of generalization ability. In contrast, the proposed PDNN scheme can imagine all cases with superior performance, except that minor flaws appear inside the scatter region with PDNN while the original PDNN avoided the flaws but at the cost of background artifacts. 
	
	For the letter profiles, since the profile geometric characters are similar with the digit-like ones, U-Net performs well and can reconstruct the scatter shape, but still insufficient in reconstruction accuracy. The PDNN scheme performs much better than U-Net and shows excellent generalization ability which can explained by the physics-driven, rather than data-driven, solving scheme. 
	
	\begin{table*}[!ht]
		\centering
		\resizebox{1.0\columnwidth}{!}{
		\caption{Relative Error of Imaging Results for the Original PDNN and the PDNN}
		\label{tab:ReEnhancedPDNN}
		\resizebox{2.0\columnwidth}{!}{
			\begin{tabular}{l|c c c c|c c c c|c c c c|c c c c}
				\toprule
				\textbf{} & \multicolumn{4}{c|}{\textbf{Digit-like}} & \multicolumn{4}{c|}{\textbf{Polygon-like}} & \multicolumn{4}{c|}{\textbf{Multiple circles}} & \multicolumn{4}{c}{\textbf{Letter}} \\ 
				\midrule
				\textbf{Indicators} & \textbf{BP} & \textbf{U-Net} & \textbf{Original PDNN} & \textbf{PDNN} & \textbf{BP} & \textbf{U-Net} & \textbf{Original PDNN} & \textbf{PDNN} & \textbf{BP} & \textbf{U-Net} & \textbf{Original PDNN} & \textbf{PDNN} & \textbf{BP} & \textbf{U-Net} & \textbf{Original PDNN} & \textbf{PDNN} \\ 
				\midrule
				Case 1   & $ 8.21\%$ & $4.20\%$ & $1.54\%$ & $1.52\%$ & $10.68\%$ & $6.55\%$ &$1.61\%$ & $1.61\%$ & $11.97\%$ & $ 8.06\%$ & $6.01\%$ & $2.51\%$ & $6.65\%$ & $4.49\%$ & $3.15\%$ & $3.03\%$          \\ 
				Case 2   & $25.64\%$ & $11.85\%$ & $7.65\%$ & $5.86\%$ & $10.76\%$ & $8.57\%$ &$11.00\%$ & $2.83\%$ & $8.13\%$ & $4.58\%$ & $2.88\%$ & $2.50\%$ & $10.22\%$ & $ 6.02\%$ & $3.25\%$ & $2.72\%$       \\ 
				Case 3   & $24.02\%$ & $9.46\%$ & $9.28\%$ & $8.09\%$ & $12.02\%$ & $8.12\%$ &$3.09\%$ & $2.68\%$ & $19.21\%$ & $16.68\%$ & $10.08\%$ & $4.20\%$ & $14.33\%$ & $9.29\%$ & $5.75\%$ & $5.62\%$       \\ 
				\bottomrule
			\end{tabular}
		}
	}
	\end{table*}
	\begin{figure*}[!ht]
		\centering
		\includegraphics[width = \linewidth]{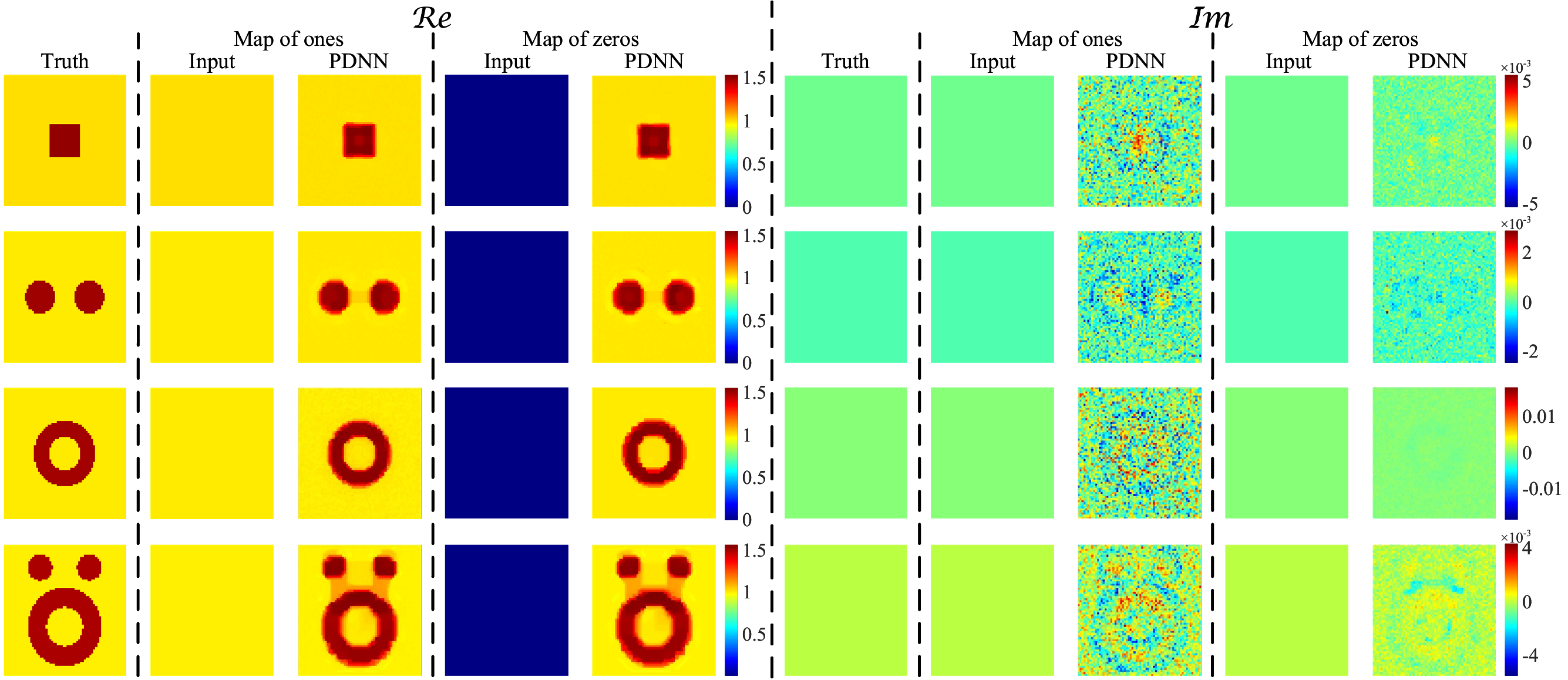}
		\caption{Tests on effects from different initial solution input to the neural network.}
		\label{fig:diffInit}
	\end{figure*}
	
	\begin{figure}[!t]
		\centering
		\includegraphics[width = 0.5\linewidth]{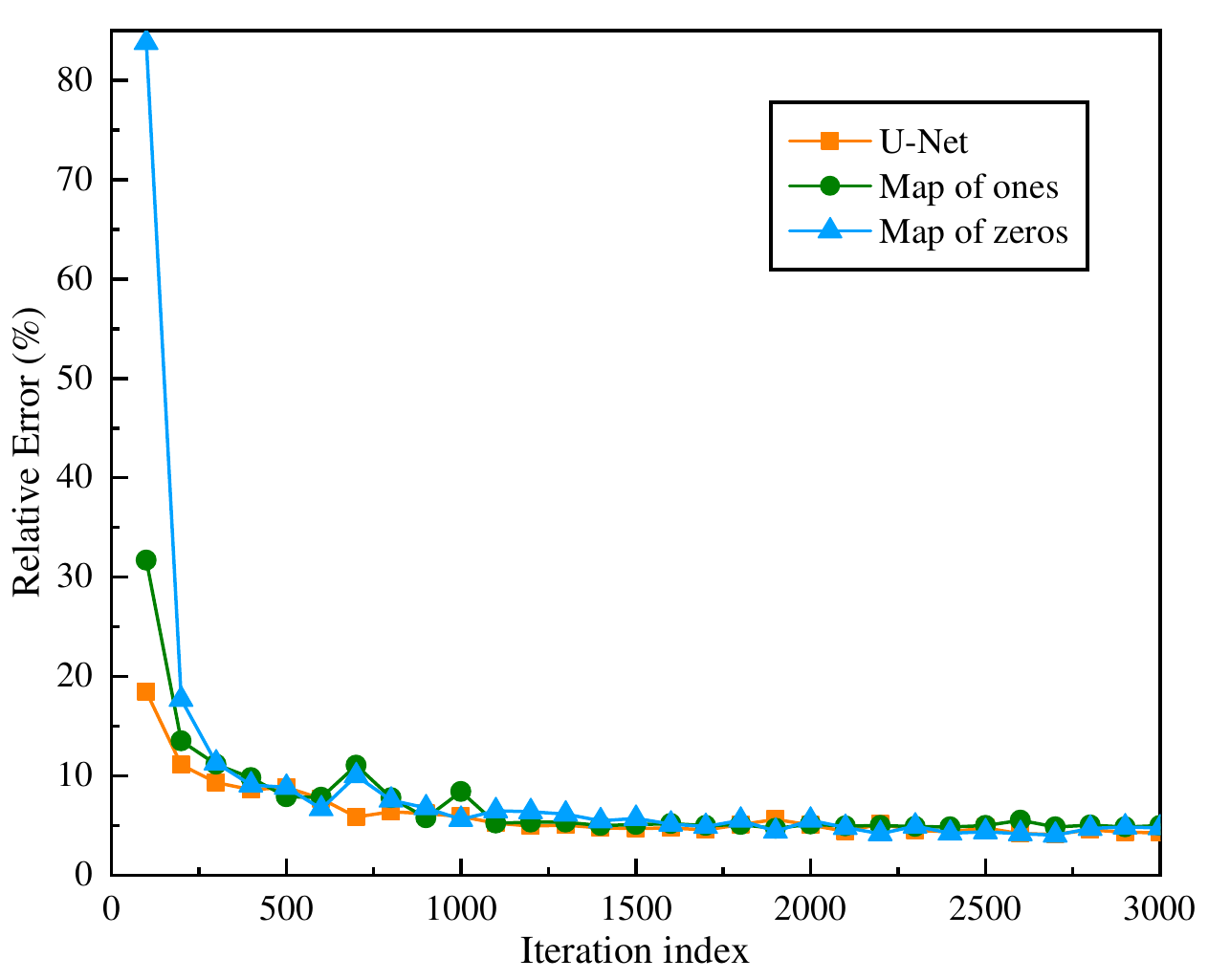}
		\caption{Effects on convergence behaviors from input initial solutions.}
		\label{fig:diffInput}
	\end{figure}
	The above conclusions are further supported by the quantitative comparison of the relative errors listed in Table\ref{tab:ReEnhancedPDNN}, where the PDNN consistently demonstrates superior accuracy. In summary, the imaging reduction procedures can significantly reduce the computational cost while keeping the superior imaging performances.
	
	\subsection{Influence of network input}
	\label{subsec:diffInit}
	
	Although the solution from U-Net is used to reduce the imaging region so that the computation cost can be lower, the proposed PDNN solver has no requirements about the initial solution which will be taken into the iterative solving scheme. To test the effects from initial solutions, the convergence rate and the obtained final solution are studied with three different initial solutions, \emph{i.e.}, U-Net solution, map of ones (free space), and map of zeros. Imaging the Austria profile, the variation of relative error as iteration index increases is shown in  Figure.~\ref{fig:diffInput}, from which convergence phenomena are observed with all three initial solutions and the convergence rate is almost same except that the prediction errors can have a large difference at the starting iteration process. While Figure.~\ref{fig:Iteration} gives the imaging results when the solution of U-Net estimator is used as the network input, Figure.~\ref{fig:diffInit} presents the imaging results with map of ones and map of zeros initial solutions. As seen, both cases yield accurate imaging results. Quantifying the reconstruction error by the defined relative error $\delta$, the resulting $\delta$ for the four examples are $0.70\%$, $1.08\%$, $1.89\%$, and $3.44\%$ with map of ones, while the value of $\delta$ being $0.67\%$, $1.11\%$, $1.59\%$, and $3.62\%$ with map of zeros. Such observations demonstrate that the imaging performance is independent from the input initial solution. 
	
	\subsection{Influence of noise}
	\label{subsec:diffSNR}
	\begin{figure*}[!t]
		\centering
		\includegraphics[width = \linewidth]{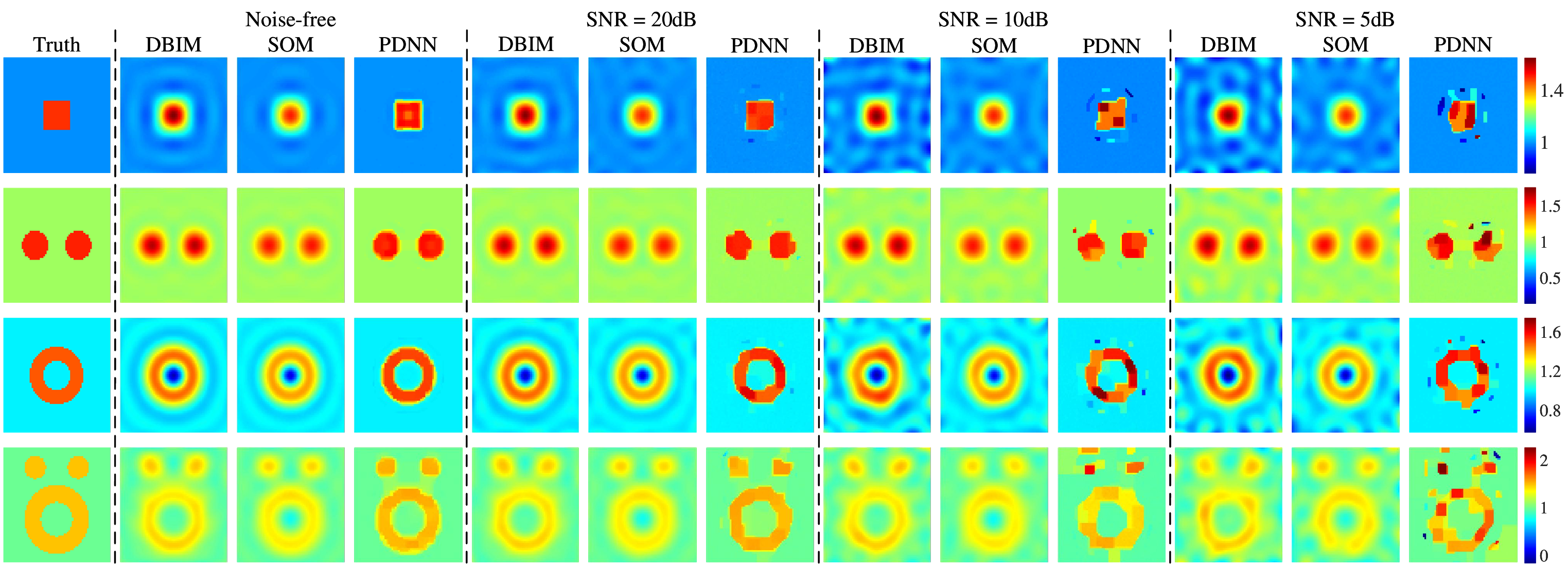}
		\caption{Tests of noise robustness of the ISP solver DBIM, SOM and PDNN.}
		\label{fig:SNR}
	\end{figure*}
	The noise robustness of the PDNN is studied by testing the imaging performance without noise and with additive Gaussian noise when SNR = 20dB, 10dB and 5dB, respectively. As shown in Figure.~\ref{fig:SNR}, DBIM and subspace-based optimization method (SOM) are robust to SNR variations that the scatter profiles can be well recovered despite slight distortion happens and background become blurred when SNR $\le$ 10 dB. Moreover, the relative permittivity at scatterer boundary regions is underestimated. With the PDNN solver, when SNR = 20dB, noise effects can be neglected, the imaging performance is significantly improved with PDNN and the scatter profiles are reconstructed at high accuracy. When SNR = 10 dB and 5 dB, the imaging performances are significantly degraded. In particular, when SNR = 5dB, strong background artifacts appear and can be wrongly identified as scatterers. 
	
	\subsection{Lossy and composite scatterers}
	\label{subsec:ComplexSca}
\begin{figure*}[!t]
		\centering
		\includegraphics[width = \linewidth]{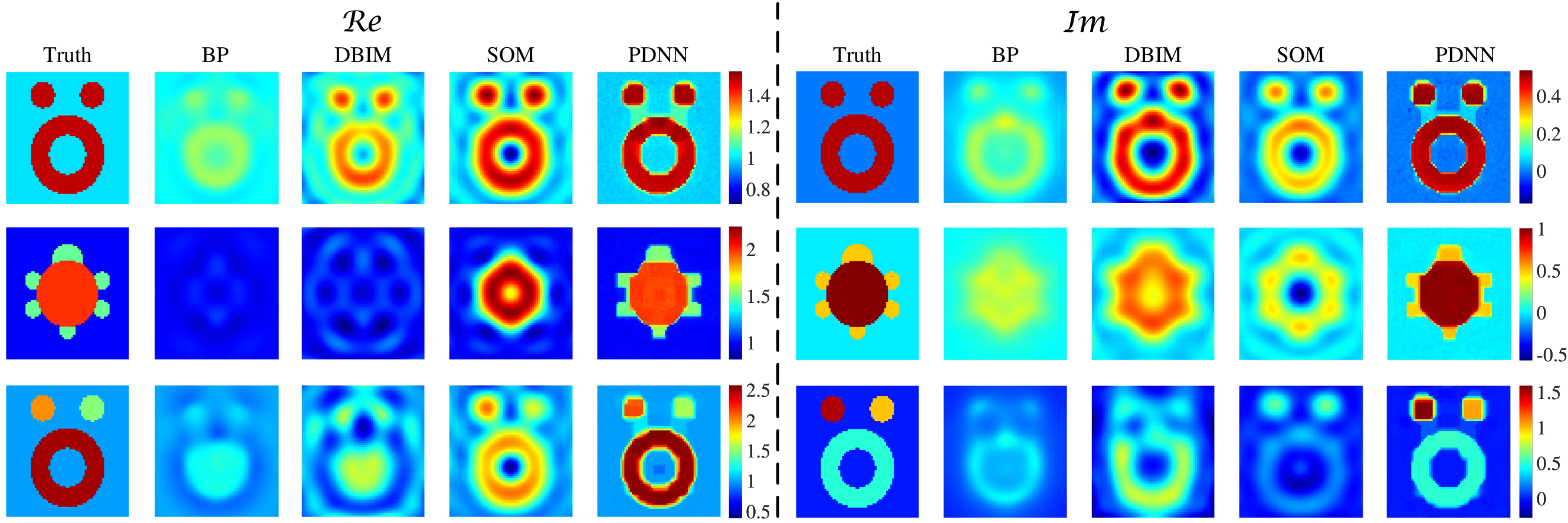}
		\caption{Imaging results of three representative lossy and composite profiles by DBIM, SOM and PDNN.}
		\label{fig:4GHzLam20LossyComposite}
	\end{figure*}

	This subsection considers more challenging and representative cases about lossy and composite profiles. Case 1: an Austria profile with $\epsilon_{r} = 1.5+j0.5$. Case 2: a turtle-like profile comprising two different dielectric materials with $\epsilon_{r} = 1.5 + j0.5$ and $\epsilon_{r} = 2 + j1.0$. Case 3: an Austria profile composed of three different dielectric materials with $\epsilon_{r} = 1.5 + j1.0$, $\epsilon_{r} = 2 + j1.5$, and $\epsilon_{r} = 2.5 + j0.5$. The relative-permittivity distribution of the three scatterers is given in  Figure.~\ref{fig:4GHzLam20LossyComposite}.
	
	As seen, DBIM can well reconstruct the shape of the homogeneous Austria profile, but the imaging performances degrade badly with piece-wise Austria profile and the turtle-like profile. SOM has improved imaging performances, with Austria profiles, the geometric information accurately recovered and the inhomogeneity can be observed although the relative-permittivity values are underestimated. However, with the turtle-like profile, which is composed a high-contrast circular scatterer surrounded by six weak and small-size scatterers, SOM obtained the image as a hexagon scatterer which is substantially different from the ground truth. In contrast, PDNN exhibits strong robustness against various types of profiles and always yields high-quality imaging results. Specifically, except for slight boundary distortion in the turtle-like profiled, both the real part and the imaginary part of the relative-permittivity distribution are accurately recovered. The observations confirm the superior capability of PDNN in handling complex composite profiles. 
	
	\subsection{Stability analysis}
	\label{subsec:Statistical}
	
	\begin{figure}[!t]
		\centering
		\includegraphics[width = 0.5\linewidth]{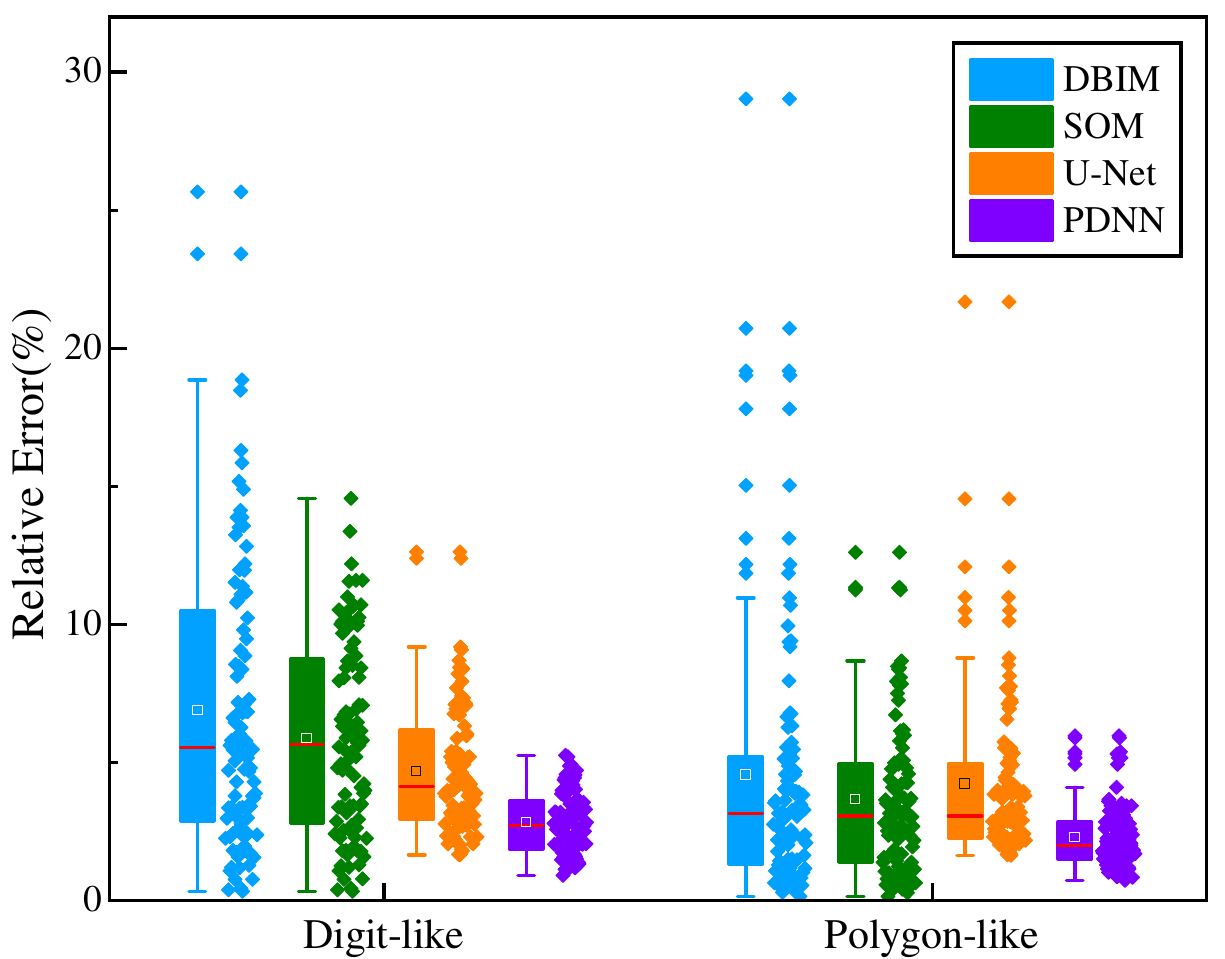}
		\caption{Boxplots of prediction errors based on 100 digit-like and polygon-like samples to indicate the stability of the imaging performance.}
		\label{fig:boxPlot}
	\end{figure}
	
	The imaging performances may vary for different scatterer profiles. Thus, a statistical analysis is conducted by imaging 100 samples of digit-like and polygon-like scatterers and comparing the variance of the corresponding prediction errors. As illustrated in Figure.~\ref{fig:boxPlot}, DBIM has a relatively high median value for relative error and the variance is also large indicating instability for different scatter samples. In contrast, SOM has improvement on the stability but is still inferior to PDNN in both median value and the variance. The U-Net solver has advantages over SOM when imaging digit-like scatterers, but the insufficient generalization ability leads to more outliers, the prediction error of which can be larger than 20\%. Therefore, the proposed PDNN behaves as the most accurate and stable solver among the compared ones.
	
	\section{Experimental validation}
	\label{sec:forExp}
	The imaging performances of PDNN are validated based on experimental data from the Fresnel Institute in Marseille, France \cite{Kamal2001Exp}. A single dielectric cylinder is imaged and the collected data are from 36 transmitters and 49 receivers which are uniformly positioned on the circle with radius 720 mm and 760 mm, respectively. 
	
	\begin{figure}[!th]
		\centering
		\includegraphics[width = .5\linewidth]{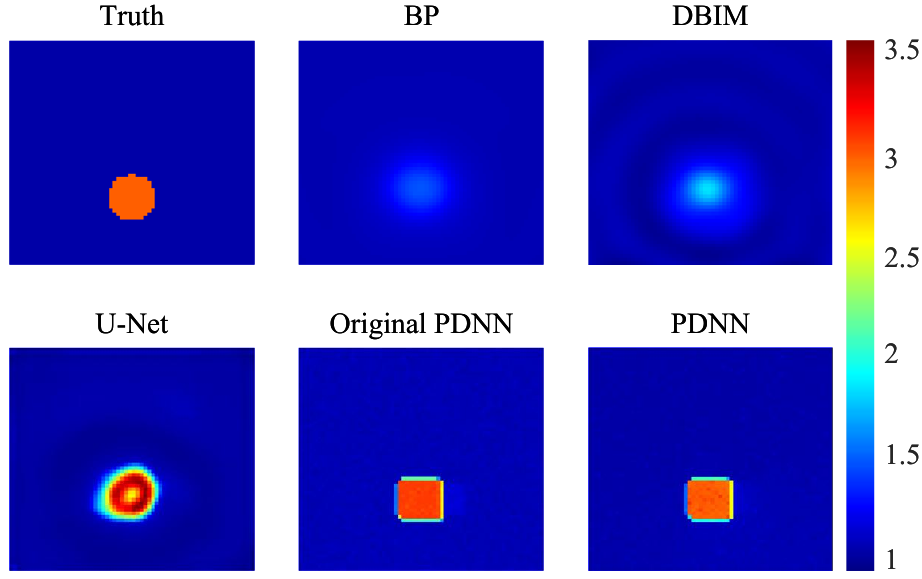}
		\caption{Imaging results of BP, DBIM, U-Net, original PDNN and PDNN based on experimental measurements.}
		\label{fig:Experiment}
	\end{figure}
	
	The imaging results based on the measurement data are presented in Figure.~\ref{fig:Experiment}. BP and DBIM severely underestimates the permittivity value. U-Net can locate and well reconstruct the shape of the circular scatterer, but fails to accurately estimate the permittivity value. PDNN, both the original one and the one reducing the imaging region, can recover the homogeneity of the scatter and the reconstructed value of relative permittivity is quite close to the ground truth. However, one also observed that with PDNN, the circular scatterer is imaged as a square one, even with similar size and the same location. Such problem maybe due to the measurement noises and is to be tackled in the future work. 
	
	\section{Conclusions}
	\label{sec:conclusions}
	This paper proposed a ISP solving scheme which applies a neural network to iteratively retrieve the contrast distribution. The training process of the neural network is not data-driven, but physics-driven through the updating of loss function in which the computation of scattered fields making use of the imaging physical laws. To reduce the computation burden, the imaging region is reduced based on the U-Net solution with morphological operations. The superior imaging performances, in terms of reconstruction accuracy and stability, are analyzed through numerical studies and confirmed with experimental results.   
	
	Specifically, PDNN established a two-channels processing mechanism for real and imaginary components to handle lossy scatterers. At each iteration, the scattered fields corresponding to the contrast distribution predicted by the neural network are computed and used to estimate the data residual term of the loss function. Incorporating the physical prior that the real part of relative permittivity of dielectric must be greater than 1, a ReLU function-based constraint term is introduced in the real-part penalty term of loss function. Additionally, Total Variation (TV) regularization is incorporated to enhance reconstruction smoothness. Based one numerical simulations, the impact from the selection of hyperparameters in loss function, the convergence testing, acceleration of the PDNN runtime, and the comparisons between BP, DBIM, SOM, U-Net and PDNN are performed. The results demonstrate the capability of the proposed PDNN scheme in imaging various types of scatterers under different noise levels and show its superiority over classical ISP solvers on reconstruction accuracy and the generalization ability for data-driven solvers. The above conclusions are confirmed based on an imaging example based on experimental data. 
	
	\bibliographystyle{unsrt}
	\bibliography{DNN}
\end{document}